\documentstyle[aas2pp4]{article}
\def \be {\begin{equation}}
\def \ee {\end{equation}}

\def \mbh {$M_{\rm BH}$}
\def \mbhb {$M_{\rm BH}$(H$\beta$)}
\def \mbulge {$M_{\rm bulge}$}
\def \lbulge {$L_{\rm bulge}$}

\begin{document}

\title{ON THE LINEARITY OF THE BLACK HOLE - BULGE MASS
RELATION IN ACTIVE AND IN NEARBY GALAXIES}
\author{Ari Laor}
\affil{Physics Department, Technion, Haifa 32000, Israel\\
laor@physics.technion.ac.il}
\begin{center}
Accepted for publication in ApJ on January 19th 2001.
\end{center}
\begin{abstract}
Analysis of PG quasar observations suggests a nonlinear 
relation between the black hole mass, \mbh, and the bulge mass, \mbulge, 
although a linear 
relation, as proposed for nearby galaxies, cannot be ruled out. 
New $M_{\rm BH}$ values for 
nearby galaxies from Gebhardt et al., and \lbulge\ measurements
for Seyfert 1 galaxies from Virani et al., are used here to obtain 
a more accurate value for the slope of the \mbh--\mbulge\ relation. The combined
sample of 40 active and non-active galaxies suggests a significantly
nonlinear relation, $M_{\rm BH}\propto M_{\rm bulge}^{1.53\pm 0.14}$.
Further support for a nonlinear relation is provided by the slope 
of the \mbh--stellar velocity dispersion relation found recently, 
and by the low \mbh\ found in late type spiral galaxies.
The mean \mbh/\mbulge\ ratio is therefore not a universal
constant, but rather drops from $\sim 0.5$\% in bright ($M_V\sim -22$) 
ellipticals,
to $\sim 0.05$\% in low luminosity ($M_V\sim -18$) bulges. Hubble Space 
Telescope determinations of \mbh\ in late type spirals, and of the bulge 
magnitude in narrow line Seyfert 1 
galaxies (both predicted to have low \mbh), can further test the
validity of the nonlinear \mbh--\mbulge\ relation.

\end{abstract}


\section{INTRODUCTION}
Massive black holes appear to be present in the cores of most or all
bulges of nearby galaxies. Magorrian et al. (1998) presented 
systematic \mbh\ determinations for a sample of 32 nearby
galaxies, and suggested that the black 
hole mass, \mbh, is proportional to the bulge mass, \mbulge, such that on 
average
\mbh/\mbulge$\simeq 0.5$\%, with a rather large scatter [$\pm 0.5$
rms scatter in log (\mbh/\mbulge) at a given \mbulge]. A lower
mean ratio of \mbh/\mbulge$\simeq 0.15\%-0.2$\% was found by
Ho (1999) and Kormendy (2000) based on compilations of \mbh\
values from various studies, again with a large scatter.
 Gebhardt et al. (2000a, hereafter G00) 
and Ferrarese \& Merritt
(2000) recently suggested there is a significantly 
tighter correlation of \mbh\ with
the bulge velocity dispersion $\sigma$, although there is some disagreement
concerning the slope of this relation (see also Merritt \& Ferrarese
2000). 

If active galactic nuclei reside in otherwise normal galaxies then their
black hole mass may also correlate with their host bulge mass. This was explored by 
Laor (1998, hereafter L98) using Hubble Space Telescope (HST) 
measurements by Bahcall et al. (1997) of the host luminosities
of a sample of Palomar-Green (PG) quasars, together with their 
\mbhb--i.e. estimates of
\mbh\ based on the size and velocity dispersion of the broad line region (BLR),
as measured with the H$\beta$ line. The PG quasars
were found to overlap well the distribution of the Magorrian et al. 
nearby galaxies in the \mbh--bulge luminosity plane, suggesting that the 
\mbh--\mbulge\ relation holds in quasars as well, and that \mbhb\ is most
likely within a factor of 2-3 of the true \mbh\ (despite potentially large
systematic errors, e.g. Krolik 2001). Further support to these results comes
from the McLure \& Dunlop (2000) study of the \mbh--\mbulge\ relation 
in a significantly larger 
sample of quasars and Seyfert galaxies.

The best fit relation for the L98 quasar sample was nonlinear, 
\mbh$\propto M_{\rm bulge}^{1.4\pm 0.4}$, but the deviation from
linearity was clearly not 
significant. The relatively large uncertainty in the slope 
was partly due to the small range of \mbh\ available 
($\sim 10^8M_{\odot}-10^9M_{\odot}$). 
However, since nearby galaxies with higher
($\sim 10^{10}M_{\odot}$) and lower ($\sim 10^7M_{\odot}$) \mbh\
also follow the quasar relation rather well (Fig.1 in L98),  
it appeared that the true \mbh--\mbulge\ relation may indeed be nonlinear.

The purpose of this paper is to better constrain the slope of the 
\mbh--\mbulge\ relation using recently published data. We use 
new \mbh\ determinations for 14 nearby galaxies reported in 
G00, and new bulge luminosity, \lbulge, determinations for nine Seyfert 1
galaxies by Virani, De Robertis, \& VanDalfsen (2000), for which 
\mbhb\ estimates can be made. In \S 2 we review the 
\mbhb\ determination in quasars, the new \mbh\ in nearby galaxies,
new \lbulge\ in Seyfert galaxies, and the resulting significantly nonlinear 
\mbh--\mbulge\ relation. In \S 3 we discuss the implications, 
provide other evidence for a nonlinear relation, and comment on the nature
of the outlying objects. The main conclusions are summarized in \S 4.

\section{THE \mbh--\mbulge\ RELATION}

\subsection{Quasars}

Figure 1a displays the \mbh\ vs. \lbulge\
distribution of 15 PG quasars
(using $H_0=80$~km~s$^{-1}$~Mpc$^{-1}$ and $\Omega_0=1$). 
The quasar sample is slightly revised
from the L98 sample with the addition of  PG~1425+267
from Kirhakos et al. (1999) at $M_V=-22.58$ and \mbh$=2.1\times 10^9M_{\odot}$, 
and the typo correction of H$\beta$ FWHM for PG~1307+085 to 5320~km~s$^{-1}$.
The quasar \mbh\ is calculated assuming the H$\beta$ line width is dominated
by gravity, and using the radius vs. luminosity relation for H$\beta$, 
which give \mbhb$=1.8\times 10^8 \Delta v_{3000}^2L_{46}^{1/2}M_{\odot}$, where
$\Delta v_{3000}={\rm H}\beta$~FWHM/3000~km~s$^{-1}$, 
$L_{46}=L_{\rm bol}/10^{46}$~erg~s$^{-1}$, and $L_{\rm bol}$ is the 
bolometric luminosity (L98).
The Spearman rank order correlation coefficient for the quasar distribution is
$r_S=0.73$, which has a probability of Pr~$=1.9\times 10^{-3}$ to occur
by chance. 
A least squares fit to the 15 PG quasars gives
\be
M_V({\rm bulge})=-8.42\pm 3.29
-(1.48\pm 0.38)\log (M_{\rm BH} /M_{\odot}) .
\ee
The rms scatter of the quasars from the above relation is 
$\Delta M_V=0.63$~mag, and $\Delta \log (M_{\rm BH}/M_{\odot})=0.43$

To convert $M_V$ to \mbulge\ we first
use the standard relation $M_V=4.83 -2.5\log (L/L_{\odot})$, 
to write the quasar relation as
\be 
\log (L_{\rm bulge}/L_{\odot})=5.3+0.59\log (M_{\rm BH}/M_{\odot}) .
\ee
The bulge luminosity is converted to mass using the Magorrian et al. fit 
(their eq. 10)
\be 
\log (M_{\rm bulge}/M_{\odot})=-1.11+1.18\log (L_{\rm bulge}/L_{\odot}) 
\ee
which, together with equation 2 gives 
\be
\log (M_{\rm BH}/M_{\rm bulge})=0.43\log (M_{\rm bulge}/M_{\odot})-7.36 .
\ee
Thus, if $M_V({\rm bulge})\propto M_{\rm BH}^{\alpha}$, then
$M_{\rm BH}\propto M_{\rm bulge}^{\beta}$, where $\beta=2.5/\alpha/1.18$,
and for the quasars we get $\beta=1.43\pm 0.37$. 

Below we describe the additional samples used to extend the study
of the \mbh--\mbulge\ relation over a larger range of \mbh.

\subsection{Nearby Galaxies}

Although the Magorrian et al. sample of nearby galaxies is large
and uniformly analyzed, we do not use it here since their \mbh\ 
values may 
have been systematically overestimated (Gebhardt et al. 2000b). 
We also prefer to avoid 
compilations of \mbh\ values from various authors, using very different 
methods, as these may
involve a range of different systematic errors which can 
increase the scatter in the \mbh--\mbulge\ relation. 

We use here the new set of \mbh\ determinations in 14 nearby galaxies 
presented by G00, all of which are based on stellar kinematics,
were made by the same set of authors, and are based on similar models, 
which should minimize the scatter
due to different systematic errors. These new \mbh\ values are based
on HST spectroscopy, and are likely to be significantly more
accurate than earlier \mbh\ values.  We add to this sample NGC~4342, 
which is part of the G00 compilation, although it was analyzed
by a different group (Cretton \& van den Bosch 1999),
and the Galaxy and NGC~4258 because of their 
exceptionally accurate \mbh\ values. 

The bulge magnitudes for all 14 nearby galaxies are taken from 
Magorrian et al. and Faber et al. (1997),
corrected for the revised distances in G00.
The bulge magnitude of NGC~4342 is obtained
using $V^0_T=12.43$ from de Vaucouleurs et al. 1991 (RC3), a bulge vs.
total magnitude difference of $0.55$ mag, based on the 
Simien \& de Vaucouleurs (1986) mean for S0 galaxies,
and a distance of 15.3~Mpc (G00). The bulge magnitude
of NGC~4258 is obtained using the water maser based distance of 7.2~Mpc
(Herrnstein et al. 1999),
$B^0_T=8.53$ from RC3, $\Delta B=2.01$ from Simien \& de Vaucouleurs,
and assuming $B-V=0.94$ for the bulge color (Fukugita, Shimasaku, \& Ichikawa 1995).
For the Galaxy we use the Bahcall \& Soneira (1980) bulge magnitude, $M_V=-18.4$,
and \mbh$=3\times 10^6M_{\odot}$ (Genzel et al. 2000). 

Figure 1a presents the \mbh\ vs. \lbulge\ distribution for the sample 
of nearby galaxies described above (see Table 1).
The least squares fit to the combined sample of PG quasars and the nearby 
galaxies gives
\be
M_V({\rm bulge})=-10.61\pm 1.23
-(1.23\pm 0.15)\log (M_{\rm BH} /M_{\odot}) ,
\ee
excluding NGC~4342 which deviates
by 3~mag from the best fit relation (see \S 3.4). The combined sample
thus gives $\beta=1.72\pm 0.21$, i.e. a significantly nonlinear relation.

A systematic relative offset between the \mbh\ estimates in quasars and
the nearby galaxies could bias the best fit slope. To explore the
possible bias introduced by such an effect we scaled down the 
quasars \mbhb\ by
a factor of 3, and refitted the data. The revised slope yielded
$\beta=1.49\pm 0.20$, which is still significantly nonlinear.
Scaling \mbhb\ down by more than a factor of three, or scaling it up
by any factor yielded a larger $\beta$. Thus, any correction for a
systematic offset in \mbh\ cannot produce a slope which is consistent with linear.

\subsection{Seyfert 1 Galaxies}

\subsubsection{The uncertainty in \lbulge\ estimates}

Ho compiled a sample of Seyfert 1 galaxies with \mbhb\ 
and bulge magnitudes based on various published
surface photometry, and noted that Seyfert 1 galaxies do not follow
the Magorrian et al. \mbh--\mbulge\ relation. They appear to have
about 5 times lower \mbh\ than nearby galaxies at a given \mbulge.
Wandel (1999) repeated this analysis using bulge magnitudes from
Whittle (1992), and noted a qualitatively similar discrepancy.

Figures 2a,b provide indirect evidence that the discrepant positions
of the Seyfert galaxies in the \mbh--\mbulge\ plane may be partly due 
to systematic errors in the bulge magnitude estimates. 
Fig.2a shows a comparison of the bulge $B$ band magnitude, 
$M_B$, for the 14 overlapping Seyfert 1 galaxies from
Ho (1999, Table 2 there), and Whittle (1992, Table 3 there, adapted to
$H_0=75$~km~s$^{-1}$~Mpc$^{-1}$ used by Ho). Fig.2b shows a similar
comparison of the seven overlapping objects in Whittle (1992) and 
Virani et al. (2000)
(both converted to $M_V$ using the mean colors of elliptical galaxies).
The different estimates deviate, sometimes systematically,
by $\sim 1-2$ magnitudes. This demonstrates the difficulty
in getting accurate \lbulge\ values in Seyfert 1 galaxies with the 
methods used by one, or more of the above authors. This difficulty probably 
arises from the relatively large distance of the Seyfert galaxies 
($\sim 100$~Mpc), compared with the nearby normal galaxies 
($\sim 10-20$~Mpc), which makes it hard to obtain an accurate
bulge + disk + point source decomposition.

Gebhardt et al. (2000b) noted that Seyfert galaxies overlap
the distribution of nearby galaxies in the $M_{\rm BH}-\sigma$ plane,
and suggested that their discrepant position 
in the \mbh--\mbulge\ plane, found earlier, is not real but is due to 
inaccurate \lbulge\ values
[rather than inaccurate \mbhb]. Nelson (2000) used
the [O III] line width as a proxy for $\sigma$ and found that nearby
galaxies, Seyfert galaxies, and quasars overlap well in the 
\mbh--$\sigma$ plane, and also concluded that there is no large systematic
bias in \mbhb. In addition, McLure \& Dunlop (2000) found direct evidence
that Seyfert galaxies fall on the same \mbh--\lbulge\ relation defined
by quasars. Below we provide additional 
direct evidence that Seyfert galaxies
follow well the \mbh--\mbulge\ relation defined by quasars and nearby 
galaxies.

\subsubsection{The Virani et al. sample}

Since Virani et al. provide detailed bulge + disk + point 
source decompositions, we suspect it is likely to be the most accurate 
of the three studies compared in Fig.2, and therefore adopt it in the 
following analysis.

Estimates of \mbhb\ could be made for nine of the 15 Seyfert 1 
galaxies in Virani et al. (listed in Table 2), as further described below. 
The bulge luminosity is calculated from the bulge 
$R_C$ magnitude measured by Virani et al., converted
to $m_V$ assuming
$V-R_C=0.61$, the mean color of  Elliptical galaxies 
(Fukugita et al., Table 3), and the distance is calculated
using the recession velocity with respect to the 3K cosmic microwave
background, $V_{3K}$, obtained from RC3 (using
$H_0=80$~km~s$^{-1}$~Mpc$^{-1}$).

The black hole mass is calculated using
$M_{\rm BH}$(H$\beta)=\Delta v^2R_{\rm BLR}/G$, where $R_{\rm BLR}$ is the 
``mean'' radius of the 
H$\beta$ emitting region in the BLR, and $\Delta v$ is the H$\beta$ FWHM.
There may be a correction factor of order unity to this expression, depending
on the BLR kinematics (e.g. Ho 1999;  McLure \& Dunlop 2000), but there is 
no accurate way to determine it yet.
 Reverberation based measurements of $R_{\rm BLR}$ are available for
six of the nine Seyferts (references listed in Table 2). 
For the other three we estimate $R_{\rm BLR}$
using $L_{\rm bol}$ and the relation 
$R_{\rm BLR}=102 L_{46}^{1/2}$ light days
(as used for the quasars above and in L98).
These three objects are: (1) Mrk~841 (PG~1501+106), for which 
$\nu f_{\nu}$(3000\AA)$=1.45\times 10^{-10}$~erg~s$^{-1}$~cm$^{-2}$, which gives
$L_{\rm bol}=2.6\times 10^{45}$~erg~s$^{-1}$ using a bolometric correction
factor $f_{\rm bol}=8.3$; (2) NGC~4253 (Mrk~766), with 
$\nu f_{\nu}$(5000\AA)$=2\times 10^{-11}$~erg~s$^{-1}$~cm$^{-2}$, 
$f_{\rm bol}=12$, giving $L_{\rm bol}=7\times 10^{43}$~erg~s$^{-1}$, 
and (3) NGC~6814, 
$\nu f_{\nu}$(5000\AA)$=3.8\times 10^{-11}$~erg~s$^{-1}$~cm$^{-2}$, 
giving $L_{\rm bol}=1.5\times 10^{43}$~erg~s$^{-1}$.
The references for the continuum flux are given in Table 2, and
$f_{\rm bol}$ is based on Fig.7 of Laor \& Draine (1993).

The six remaining objects in Virani et al. were not used here for
the following reasons: Mrk~231 is strongly 
interacting and its spectrum is significantly absorbed, making it very
difficult to get a reliable estimate of its \mbhb\ and
\lbulge; Mrk~789 was reclasiffied as
a starburst by Osterbrock \& Martel (1993); NGC~3080 has no publicly
available optical spectrum; NGC~3718 displays a very red LINER spectrum 
without H$\beta$
(Barth, Filippenko, \& Moran 1999);
and NGC~4235 and NGC~5940 have optical spectra in Morris \& Ward 
(1988), but of a very low S/N in the H$\beta$ region.

Figure 1b shows the positions of the Virani et al. Seyfert galaxies
in the \mbh--\lbulge\ plane, and demonstrates they follow the quasar 
\mbh--\lbulge\ relation well. This provides a direct indication that 
Seyfert galaxies are not significantly offset from nearby galaxies, 
as suggested in earlier studies.

The least squares fit to the combined sample of  
PG quasars and Seyfert galaxies gives
\be
M_V({\rm bulge})=-7.73\pm 1.42
-(1.56\pm 0.17)\log (M_{\rm BH} /M_{\odot}) ,
\ee
which, as seen in Fig.1b, is very close to the fit for the quasars 
alone. However, the error
on the slope is significantly reduced due to the 
increased range of \mbh\ in the combined sample. The combined sample
gives $\beta=1.36\pm 0.15$, again a nonlinear relation, though not as
steep as the one found
for the combined sample of quasars and nearby galaxies.

\subsection{All Objects}

A least squares fit to the combined sample of 40 objects 
(15 PG quasars, 16 normal galaxies, and 9 Seyfert galaxies) gives
\be
M_V({\rm bulge})=-9.33\pm 1.08
-(1.38\pm 0.13)\log (M_{\rm BH} /M_{\odot}) ,
\ee
with $r_S=-0.80$, Pr~$=4\times 10^{-10}$, 
$\Delta M_V=0.64$~mag, and $\Delta \log (M_{\rm BH}/M_{\odot})=0.46$.
The best fit slope of $-(1.38\pm 0.13)$ implies that
$M_{\rm BH}\propto M_{\rm bulge}^{1.54\pm 0.15}$, suggesting
that the \mbh--\mbulge\ relation in active and non-active galaxies
is significantly nonlinear.

\section{DISCUSSION}

\subsection{The Implication of a Nonlinear Relation}

The nonlinear \mbh--\mbulge\ relation implies that
\mbh/\mbulge\ is not universal, but rather increases with bulge luminosity.
Following the derivation in Eqs.2-4 for the combined sample fit 
(eq.7) gives
\be
\log (M_{\rm BH}/M_{\rm bulge})=0.54\log (M_{\rm bulge}/M_{\odot})-8.56 .
\ee
Thus, in bright ellipticals, say at $M_V=-22$ [i.e. 
$\log (L_{\rm bulge}/L_{\odot})=10.73$, and thus 
$\log (M_{\rm bulge}/M_{\odot})=11.55$]
the nonlinear relation gives \mbh/\mbulge$\sim 0.5$\%, as for example
observed in M~87 (Magorrian et al.). But, in the bulges
of late type spirals, say at $M_V=-18$ 
[i.e. $\log (L_{\rm bulge}/L_{\odot})=9.13$, and
$\log (M_{\rm bulge}/M_{\odot})=9.67$],
this ratio drops to \mbh/\mbulge$\sim 0.05$\%, as 
observed in the Galaxy and in NGC~7457 (the two leftmost points in
Fig.1a). This low \mbh/\mbulge\ is consistent with the recent upper
limit of \mbh/\mbulge$\le 0.07$\% in NGC~4203, a LINER S0 galaxy with
$M_V=-18.1$ (Shields et al. 2000), and with the general result of Salucci et al. 
(2000) that \mbh$\le 10^6-10^7 M_{\odot}$ in late type spirals. Based on 
a few detections, Salucci et al. also commented that spirals
appear to follow a steeper than linear \mbh--\mbulge\ relation (their Fig.8), but
they do not provide details which would allow a quantitative comparison.

\subsection{Additional Evidence for a Nonlinear relation}

Ferrarese \& Merritt found that $M_{\rm BH}\propto\sigma^{4.8}$ and suggested
that since $L_{\rm bulge}\propto\sigma^4$ and $M_{\rm bulge}\propto 
L_{\rm bulge}^{1.2}$, then 
$M_{\rm bulge}\propto\sigma^{4.8}$ which implies a linear relation 
$M_{\rm BH}\propto M_{\rm bulge}$. 
However, the projection of the core fundamental plane on the 
$L_{\rm bulge}-\sigma$ plane (the ``Faber-Jackson'' relation) is much 
flatter than assumed by Ferrarese \& Merritt.
For example, Nelson \& Whittle (1996) find 
$L_{\rm bulge}\propto\sigma^{2.7\pm 0.3}$ in Seyfert galaxies, and
inspection of the G00 sample suggests a similar relation  
(see also Fig.4b in Faber et al.). The Ferrarese \& Merritt  
relation then implies $M_{\rm BH}\propto M_{\rm bulge}^{1.78}$, while
the G00 relation, $M_{\rm BH}\propto\sigma^{3.75}$, implies
$M_{\rm BH}\propto M_{\rm bulge}^{1.39}$, both of which are comparable
to the nonlinear slopes found here.

\subsection{Comparison with the McLure \& Dunlop Best Fit Slope}

McLure \& Dunlop suggest a linear \mbh--\mbulge\ 
relation for their sample of active galaxies, with \mbh/\mbulge=0.25\%. 
Specifically, they find a slope of $-0.61\pm 0.08$ in the
bulge magnitude vs. log~\mbh\ plane, which corresponds to a slope of
$-1.64\pm 0.22$ [$=1/(-0.61\pm 0.08)$] in log~\mbh\ vs. bulge magnitude,
consistent with the slope of $-1.56\pm 0.17$ found here for
the combined sample of quasars and Seyfert galaxies (\S 2.3.2). However, 
they then assume \mbulge$\propto$\lbulge$^{1.31}$, as infered by Jorgensen, 
Franx \& Kjaergaard
(1996) from their Gunn-r study of elliptical galaxies in nearby rich clusters.
This slope then leads to $M_{\rm BH}\propto M_{\rm bulge}^{1.16\pm 0.16}$,
consistent with a linear relation. However, the Magorrian et al. dependence,
\mbulge$\propto$\lbulge$^{1.18\pm 0.03}$, is more appropriate
here since it was determined directly by modeling the stellar dynamics,
rather than through scaling relations (Dressler et al. 1987), it is deduced 
in the $V$ band,
consistent with our use of $M_V$, and it is based on a sample of nearby
galaxies which partly overlaps the G00 sample.

\subsection{The Nature of the Outliers}

The large scatter in the \mbh--\lbulge\ relation was noted by many aothors.
Below we suggest that this may be
due to relatively few anomalous galaxies.
For example, NGC~4342 deviates significantly 
from the \mbh--\lbulge\ relation,
being 3 magnitudes fainter than expected for its \mbh\ (Fig.1a).
Similarly, NGC~4486B with
$\log (M_{\rm BH}/M_{\odot})=8.96$ and $M_V=-17.57$
(Magorrian et al.), is 
4 magnitudes fainter than expected (see also Fig.1 in L98).
Faber et al.  noted that NGC~4486B deviates
significantly from the core fundamental plane relation, being 
4 magnitudes fainter than expected for its central velocity dispersion
of $\sigma=200$~km~s$^{-1}$ at $M_V=-17.57$ (their Fig.4b).
NGC~4342, with $\sigma=225$~km~s$^{-1}$ and $M_V=-17.94$, is 
also similarly offset from the core fundamental plane. 
Faber (1973) suggested that these
offset galaxies
are ``tightly bound cores of normal elliptical galaxies whose outer
regions have been stripped away in tidal interactions with more massive
companions''. These anomalous galaxies would then have a $\sigma$ which
corresponds to their original \mbulge, but a lower current \lbulge.
This may explain why they are outliers in the \mbh--\lbulge\ relation,
but not in the \mbh--$\sigma$ relation, as found by G00 
and Ferrarese \& Merritt. Exclusion of galaxies which do not follow
the core fundamental plane may significantly reduce the scatter
in the \mbh--\lbulge\ relation.

\section{CONCLUSIONS}

The main aim of this paper is to point out that the \mbh--\mbulge\
relation appears to be nonlinear, with \mbh/\mbulge\ increasing
from $\sim 0.05$\% in the least luminous bulges with detected black holes
($M_V\sim -18$) to $\sim 0.5$\% in bright ellipticals ($M_V\sim -22$).

The slopes of the \mbh--$\sigma$ relation found by G00 and
Ferrarese \& Merritt, together with the Faber-Jackson relation, also indicate
a nonlinear \mbh--\mbulge\ relation, as do the low upper limits on \mbh\ in
late type spirals. The larger scatter in the \mbh--\lbulge\ relation,
compared with the \mbh--$\sigma$ relation, 
may be due to relatively few anomalous galaxies which do not follow the core 
fundamental
plane relation. 

Forthcoming systematic HST determinations of \mbh\ in large samples of nearby 
galaxies (e.g. Marconi et al. 2000), and \lbulge\ determinations in large 
samples of active galaxies, will allow to 
establish the slope, scatter, and nature of outliers in the
\mbh--\lbulge\ relation. Late type spirals, and narrow line Seyfert 1
galaxies (e.g. Laor 2000) would be especially
important since they are expected to have particularly low \mbh, and thus provide 
the strongest 
leverage on the strength and slope of the \mbh--\lbulge\ relation.
 
\acknowledgments

Some very useful comments by Luis Ho, Charlie Nelson, and the referee 
are greatly appreciated. The NASA Extragalactic Database (NED) was 
extensively used and is gratefully acknowledged.

\small

\begin{table}
\caption{GEBHARDT ET AL. SAMPLE} 
\begin{tabular}{lcclcc}
\tableline \tableline 
Galaxy &  $M_V$ & $M_{\rm BH}^a$ & Galaxy &  $M_V$ & $M_{\rm BH}^a$ \\ 
\tableline
N821  & $-$21.10 & 7.70 & N4342 & $-$17.94 & 8.48 \\
N1023 & $-$20.38 & 7.59 & N4473 & $-$20.79 & 8.00 \\
N2778 & $-$19.50 & 7.30 & N4564 & $-$19.90 & 7.76 \\
N3377 & $-$19.97 & 8.00 & N4649 & $-$22.34 & 9.30 \\
N3379 & $-$20.66 & 8.00 & N4697 & $-$21.26 & 8.08 \\
N3384 & $-$19.87 & 7.26 & N5845 & $-$19.69 & 8.51 \\
N3608 & $-$21.11 & 8.04 & N7457 & $-$18.51 & 6.53 \\
N4291 & $-$20.66 & 8.18 & N4258 & $-$19.69 & 7.62 \\
\tableline 
\normalsize
\end{tabular}
$^a$ Log \mbh/$M_{\odot}$\\
\end{table}

\begin{table}
\caption{VIRANI ET AL. SAMPLE} 
\begin{tabular}{lrccrcc}
\tableline \tableline 
Galaxy & cz$^a$ & $M_V$ & H$\beta^b$ & $R_{\rm BLR}^c$ & $M_{\rm BH}^d$ & Ref.$^e$\\ 
\tableline
Mrk 817 & 9481 & $-$19.48 & 4490 & 15.0 & 7.77 & 1, 6\\ 
Mrk 841 &10852 & $-$20.61 & 5470 & 52$^f$ & 8.49 & 7, 3\\ 
N3227   & 1472 & $-$18.40 & 4920 & 10.9 & 7.71 & 1, 6\\
N3516   & 2696 & $-$20.51 & 4950 & 7 & 7.53 & 2, 5$^g$\\  
N4051   &  945 & $-$17.42 & 1170 & 6.5 & 6.24 & 1, 6\\ 
N4151   & 1238 & $-$19.00 & 5910 & 3.0 & 7.31 & 1, 6\\ 
N4253   & 4101 & $-$19.22 &2400 & 8.5$^f$ & 7.00 & 4, 4\\
N5548   & 5354 & $-$20.63 &6300 & 21.2 & 8.21 & 1, 6\\
N6814   & 1359 & $-$18.28 &3950 & 4.0$^f$ & 7.08 & 5, 5$^g$\\
\tableline 
\normalsize
\end{tabular}
$^a$ in km~s$^{-1}$ ($V_{\rm 3K}$ from RC3); 
$^b$ FWHM in km~s$^{-1}$;
$^c$ measured in light days, based on reverberation mappings ;
$^d$ Log \mbh/$M_{\odot}$;
$^e$ References, first number is for $R_{\rm BLR}$, second is for 
H$\beta$ FWHM;
$^f$ estimated from continuum luminosity;
$^g$ Mean of the high and low H$\beta$ FWHM values.\\
References: 1. Kaspi et al. (2000); 2. Wanders et al. (1993); 3. Boroson \& Green
(1992); 4. Osterbrock \& Pogge (1985); 5. Rosenblatt et al. (1992);
6. Wandel, Peterson \& Malkan (1999); 7. Neugebauer et al. (1987)
\end{table}
\onecolumn

\newpage
\begin{figure}
\plotone{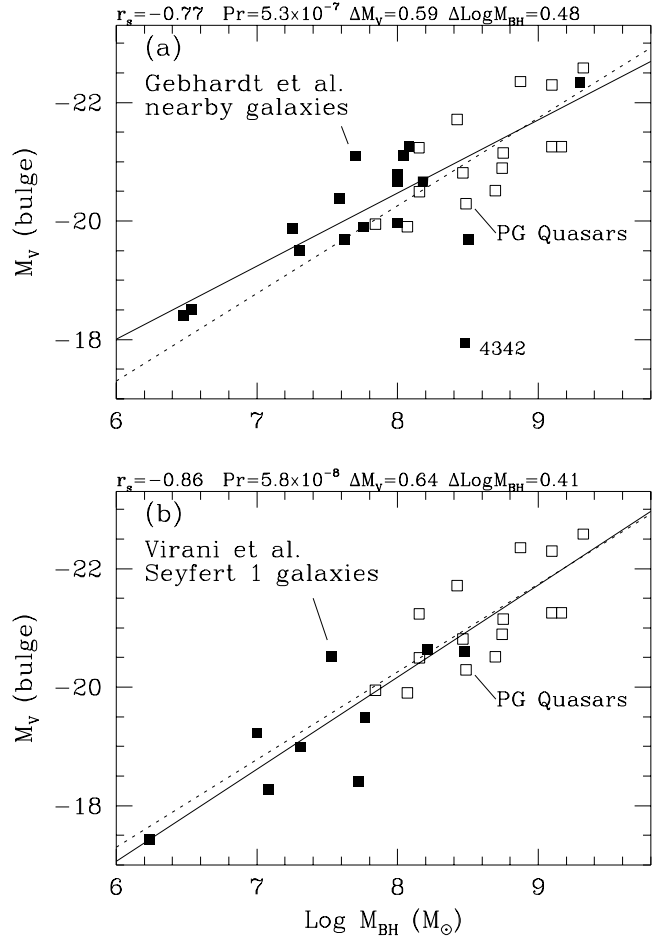}
\caption{The \lbulge--\mbh\ correlations.
Open squares mark the 15 PG quasars, filled squares mark other objects, the
dashed line is a least squares fit to the quasars only, and the 
solid line is a least squares fit to the combined sample of each panel. 
(a): The 14 nearby galaxies from G00 + the Galaxy, NGC~4342, and NGC~4258. (b):
The nine Seyfert 1 galaxies from Virani et al. 
The Spearman rank order correlation coefficient, its significance
level, and the rms scatter are indicated above each panel. 
In both samples the
best fit slope indicates a significantly 
nonlinear \mbh--\mbulge\ relation.}
\end{figure}

\newpage
\begin{figure}
\plotone{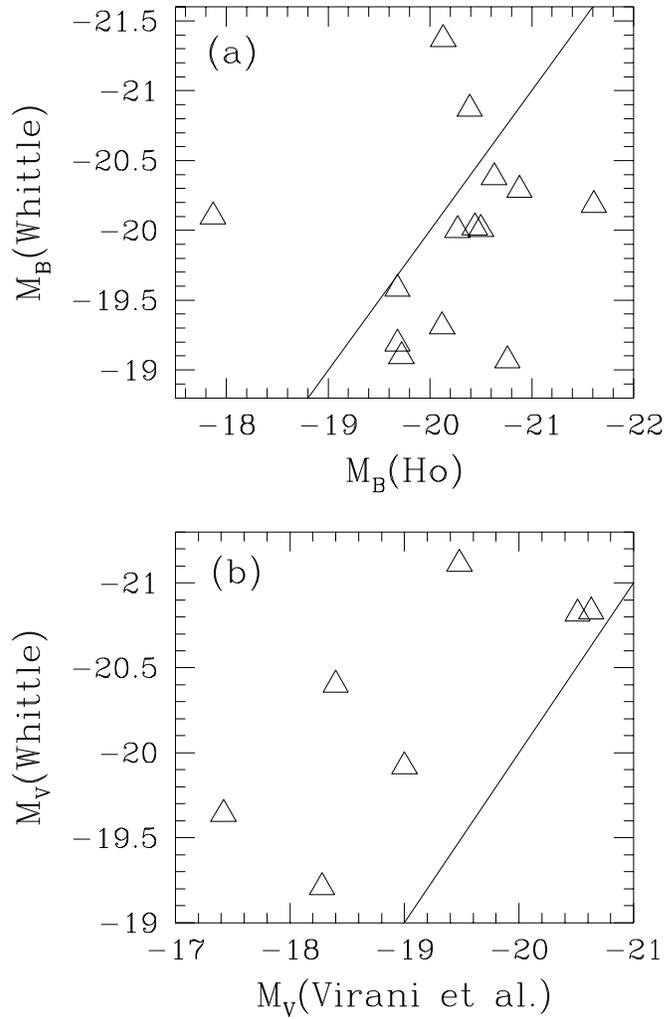}
\caption{Comparisons of different absolute bulge magnitude estimates
for Seyfert 1 galaxies. The solid
line represents identical magnitudes.
(a): The overlapping galaxies in the samples
of Ho and Whittle. (b): Overlapping
galaxies from Virani et al. and Whittle.
Note the large and systematic scatter in both panels.}
\end{figure}

\end{document}